\documentclass[letterpaper,12pt,final,3p,times,longbibliography]{elsarticle}
\pdfoutput=1
\usepackage{graphicx,epsfig,url,lineno}
\usepackage{amssymb,fontenc,times,mathptmx}
\usepackage{wrapfig,rotating}
\usepackage[usenames]{color}
\usepackage{subcaption}
\usepackage{siunitx}
\usepackage[colorlinks = true,
            linkcolor = black,
            urlcolor  = blue,
            citecolor = blue,
            anchorcolor = blue]{hyperref}
\usepackage{tikz}
\usepackage{amsfonts}
\usepackage{systeme}

\usepackage{enumitem}
\usepackage{datatool-base}

\journal{LCWS 2023 Proceedings}

\def \ee {\mathrm{e}^{+}\mathrm{e}^{-}}

\def \mumu {\mu^+ \mu^-}

\def \eemmg { \mathrm{e}^{+}\mathrm{e}^{-} \rightarrow \mu^+ \mu^- (\gamma)}

\def \invfb {\mathrm{fb}^{-1}}

\def \sqrtsLS {\sqrt{s}_{\mathrm{LS}}}
\def \sqrtsGEN {\sqrt{s}_{\mathrm{GEN}}}

\def \g1z {g_{1}^{\mathrm{Z}} }

\def \p12mag {|\vec{p}_{12}|}
\def \xp {x^{\prime}}

\def \com#1 {\textcolor{red}{#1}}

\begin{document}

\begin{frontmatter}

\title{{\Large{\bf 
Using the GP2X framework for  
center-of-mass energy precision studies  
at $\ee$ Higgs factories}}}

\author{Brendon Madison}
\address{Department of Physics and Astronomy, University of Kansas, \\
  Lawrence, KS 66045, USA}

\pdfoutput=1
\begin{abstract}
Two channels for measuring the absolute 
center-of-mass energy, $\sqrt{s}$, and collision beam energies, $E_-$ and $E_+$, are investigated for future $\ee$ Higgs factories. These two channels of DiMuons and Bhabhas are simulated in event generators and a new fast Monte Carlo, GP2X, has been developed to boost the events into the lab frame, and thus include luminosity spectrum effects from beamstrahlung and beam energy spread. GP2X also includes 
tracker and ECAL detector resolution effects. 
The performance of GP2X with WHIZARD, as GP2WHIZ, is found to be statistically consistent with iLCSoft while GP2X with KKMC, as GP2KKMC, is consistent within 3\%. We use the design concept for the ILC at $\sqrt{s}=250$~GeV and ILD. Taking advantage of ILD's high-precision tracker we find precision near the 1-10 MeV level for all of $\sqrt{s}$ , $E_{\pm}$ for a 
100~$\invfb$ dataset. This is done using a new six parameter Beta distribution convolved with Gaussian fit. A Fourier transform deconvolution method with Savitzky-Golay 
filtering is used to improve the fitting of detector level data. Feasibility of these measurements will depend on future studies and detector calibration.
\end{abstract}

\end{frontmatter}

\begin{center}
{\it Talk presented at the International Workshop on Future Linear Colliders (LCWS 2023),\\
15-19 May 2023. C23-05-15.3}
\end{center}

\pdfoutput=1
\section{Introduction}



Various $\ee$ collider concepts are under investigation. While they vary in 
maturity, cost, and potential location, all of the 
discussed linear colliders, ILC, CLIC, $\mathrm{C}^{3}$, HALHF, 
HELEN, and ReLiC, and circular colliders like CEPC and FCC-ee, are relevant to this study assuming a detector equivalent to ILD is deployed. 
For the sake of simplicity and maturity we have chosen to use the ILC at $\sqrt{s}=250$~GeV (ILC250) and ILD design concepts for this study \cite{ILD}. Particularly the ILD integrated tracker system, which allows for precise tracking of charged particle momenta, 
especially at wide angle, and for muons that do not experience appreciable bremsstrahlung energy loss 
in the detector material.
Still, given the larger cross-section Bhabha scattering has considerable statistics at wide angles. 
The theory, measurement and simulation of which was thoroughly investigated at LEP \cite{Beenakker} \cite{AcciarriBhabha}.

In this paper we expand on previous work of using DiMuons and Bhabhas for energy precision of the center-of-mass, $\sqrt{s}$ \cite{BMadison}. Initial work here was done using iLCSoft, a collection of software that is used primarily by the ILC and CLIC collaborations \cite{ILD} \cite{ilcsoft}. This initial work was then repeated, to ensure robustness, using other event generators and a new Monte Carlo. In particular three event generators of KKMC, BHWIDE and WHIZARD were used to ensure compatible results \cite{KKMC} \cite{BHWIDE} \cite{WHIZARD}. KKMC and BHWIDE are LEP-era event generators that utilize Yennie-Frautschi-Suura (YFS) exponentiation to do near infinite log order in radiative corrections. KKMC is used primarily for dilepton events that follow annihilation processes. Since Bhabha scattering is a scattering process it is instead handled in BHWIDE, as BHWIDE is used for wide angle Bhabha scattering. Neither KKMC or BHWIDE have fully verified handling for polarization. To get polarization dependence a newer event generator, WHIZARD, is used. WHIZARD is a general purpose high energy physics event generator. WHIZARD is also used in iLCSoft. WHIZARD does not use YFS and, as such, higher order radiative corrections have to be handled using WHIZARD's ISR protocols and adding in the higher order processes manually and having WHIZARD compute their matrix elements.

A new Monte Carlo, GP2X, is used to convolve beam energy spread, beamstrahlung, the beam boost, and ILD tracker response, with the event generators. As such our simulated data, and its analysis, is inclusive to numerous effects that are expected in future experimental data. The approach of GP2X can be summarized as an efficiently sorted and automated high energy physics library. Which has been made possible by a new algorithm known as Fitted Orthogonal Sampling (FOS). With minimal user input GP2X can determine the nuance parameters of a given event generator dataset and beam interaction dataset and then convolve the two datasets into a generator level dataset and detector level dataset.

We assume similarity of performance for techniques used for $\mumu$ and $\ee$ in terms of luminosity and energy precision. We investigate using these channels to infer the collision beam energies, written as $E_-^C$ and $E_+^C$. This is done using an algebraic method as outlined in similar work \cite{Wilson}. For fitting of these distributions we initially repeat the methods of previous work \cite{BMadison} but later introduce a new deconvolution method building on other recent work \cite{Wilson}. The issues of the previous method, the non-integrable singularities and the presence of a non-physical delta function as found in various implementations of CIRCE \cite{Ohl:1996fi}, are avoided. The fitting of detector level data is also handled with plausible quality. We use a goodness of fit, via $\chi^2/NDoF$, and overall precision to evaluate the fitting methods and simulation methods against each other. We find that GP2X, within this scope and using WHIZARD, is statistically consistent with iLCSoft. Finally, we summarize and provide possible next directions for the work done here.

\pdfoutput=1
\section{GUINEAPIG To X (GP2X)}

\begin{figure}[!h]
\centering
\includegraphics[width=0.95\textwidth]{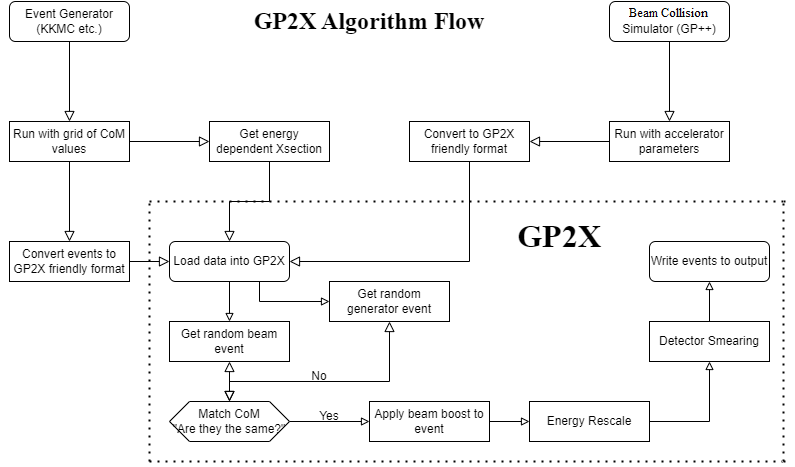}
\caption[]{\small \sl 
Algorithm flow diagram that, to some extent, follows the typical notation of algorithm flow diagrams as used in computer science. 
}
\label{fig:alg}
\end{figure}

Generating events for collider experiments, particularly when doing precision studies, requires a precise simulation of the beam collision, the particle interaction, and the detector measurements. Here we present GP2X as a software framework for the handling of such simulation \cite{GP2X}. For $\ee$ colliders GUINEAPIG is used for beam collision simulation \cite{GP}. GUINEAPIG is used in GP2X and is the GP of GP2X. We use X to refer to any event generator, which handles the particle interactions. Particular to this study X would be any of KKMC, WHIZARD and BHWIDE \cite{KKMC} \cite{WHIZARD} \cite{BHWIDE}. Detector simulation was not done in a full simulation such as in GEANT4. Instead it was simplified into a parametric model for stochastic smearing of the track $p_T$ for tracker measurements and stochastic Gaussian smearing of the energy for calorimeter measurements. The details of this model can be found in other work \cite{Wilson}.
We provide an algorithm flow diagram for GP2X in Figure \ref{fig:alg}.

Initial state simulation is handled by GUINEAPIG++ (GP++) where beam collision effects, such as beamstrahlung, are computed. GP++ uses 
input cards for defining the accelerator configuration and computational parameters and optional beam particle files. 
These were taken from other work done with ILC250 in order to ensure that the GP++ performance here is similar to other ILC250 work. The output of GP++ is unique to GP++ so it is converted to a ROOT format \cite{ROOT}. In parallel to this, at least in terms of job pipelining, an event generator is run with a grid of 100 MeV in different $\sqrt{s}$ energies. These are then combined into a ROOT file. The reason for this gridding is that we will need to be able to sample the entire luminosity spectrum provided by GP++. As such we shouldn't match a 250 GeV event with an initial electron and positron with center of mass energy of 240 GeV.

To match events between the event generator dataset and the GP++ dataset according to their $\sqrt{s}$ we may choose to randomly choose events and then veto ones that do not match. This solution is quite slow being of $\mathcal{O}(a^2 N^2)$ for matching two datasets. There are other algorithms for doing this more efficiently such as Latin Hypercube and Orthogonal Sampling \cite{MCBOOK}. In particular we look at Orthogonal Sampling where sampling is broken into orthogonal neighborhoods. Thus a random variable in one dimension is already matched to 
a neighborhood in a second, or more, dimension(s). As we already have our event generator $\sqrt{s}$ gridded we now only need a way to convert these grids to their corresponding GP++ matches. To do this we do a fit using the entry number in the dataset for the event generator dataset. From this fit we can compute the range of entry numbers in the event generator that match a given $\sqrt{s}$ randomly sampled from GP++. As such we call this method Fitted Orthogonal Sampling (FOS). Due to the algorithmic form of FOS we postulated that it should go as $\mathcal{O}(aN)$, where $a$ is a scalar constant and $N$ is the number of operations. This was confirmed using timing tests wherein it was determined that $a\approx 4$ here. As such, using FOS increased the match throughput from roughly $\frac{100k}{week}$ to roughly $\frac{300k}{hr}$.

With this we have matched an initial state electron-positron pair from the luminosity spectrum with center of mass 
energy, $\sqrtsLS$, to final state particles from the event generator file with a different 
center-of-mass energy $\sqrtsGEN$. 
The four vectors for the final state particles are then adjusted to ensure that they have the same 
center-of-mass energy as the initial state electron-positron pair, namely $\sqrtsLS$.
We choose a rescaling procedure that retains zero net momentum in the 
original event generator rest frame\footnote{There is no unique choice, but this has the benefit of relative ease}.
We therefore rescale the momenta magnitudes using a single scalar $K$; $K$ rescales all momenta magnitudes by the same factor while keeping the particle directions unchanged.
The value of $K$ is found by solving for the root of Eqn.~\ref{eqn:Rescaling} 
using the Newton-Raphson method:
\begin{equation}
f(K) = \sum_{i=1}^{N_{\mathrm{GEN}}}\sqrt{K^2 p_i^2 + m_i^2} - \sqrtsLS = 0 \: \: .
\label{eqn:Rescaling}
\end{equation}
Here $p_{i}$ is the momentum magnitude and $m_{i}$ the mass for each of the $N_{\mathrm{GEN}}$ generator particles. 
The initial estimate of $K$ denoted $K_{0}$ is computed from 
\begin{equation}
K_0  = \left(\sqrtsLS - \sum_{i=1}^{N_{\mathrm{GEN}}} \frac{m_i^2}{E_i}\right) / \left(\sqrtsGEN - \sum_{i=1}^{N_{\mathrm{GEN}}} \frac{m_i^2}{E_i}\right) \: \: ,
\end{equation}
where $E_{i}$ is the generator particle energy.
Two iterations are found to be sufficient for sub-ppm precision.
The final state particles are then boosted using the boost values of the initial particle pair so that they are in the lab frame.

With rescaling done, the final state particles are now 
suitably adjusted generator level data. 
To approximate detector level data a parametric model was used for tracker $p_{T}$ smearing and 
$18\%/\sqrt{E(\mathrm{GeV})} \oplus 1\%$ Gaussian smearing 
to model approximately ECAL energy measurements \cite{Wilson}. With this we then have detector level data and the GP2X program finishes by writing the data to a ROOT TTree.

\section{GP2X Validation}\label{sec:valid}

For validation we first communicate the values used and choices made here. The binning in energy for event generator $\sqrt{s}$ values was 0.1 GeV. Each energy binning was simulated for 10000 events. A secondary program was written to ensure that the random seeding was independent for KKMC and BHWIDE as the default seeding does not guarantee this. A test was conducted with different seeding and 0.05 GeV binning and we found no statistically significant difference in simulation precision while there was a statistically significant increase in simulation time. Both DiMuon and Bhabha event generation was cut to be below 175 degrees and above 5 degrees in $\theta$. It was observed that, for Bhabha event generation, one needs more than 10000 events to get a decent sample of wide angle Bhabhas coming from electroweak contributions. Since the ratio of cross-sections is order of 100. This was not done here as the simulation time becomes unfeasible. Thus, future studies will need to devise ways to handle larger event generation and sampling. Initial attempts into this using parallel computing and integrating parts of GP2X and GUINEAPIG++ into WHIZARD are being developed and investigated.

In order to ensure proper operation the output of GP2X was checked in a variety of ways. First, the output kinematics were checked with expected values given the original accelerator kinematics. The boost along the beam axis, and its spread, were found to be consistent with the original beam. The inferred collision beam energies, as shown in greater detail later, also confirmed the original beam energy spreads. Lastly GP2X was checked against iLCSoft at generator and detector level using DiMuons and $\sqrt{s_{p}}$~ as the measure. The purpose being that this check would verify both kinematic performance and GP2X's modeling of ILD's tracker. This was checked using GP2X in two configurations. First using unpolarized DiMuons from KKMC, then using an even mix of (+0.8,-0.3) and (-0.8,+0.3) sets of DiMuons from WHIZARD. For both configurations of GP2X the corresponding events from iLCSoft were used. 

\begin{figure}[!h]
\centering
\includegraphics[width=0.75\textwidth]{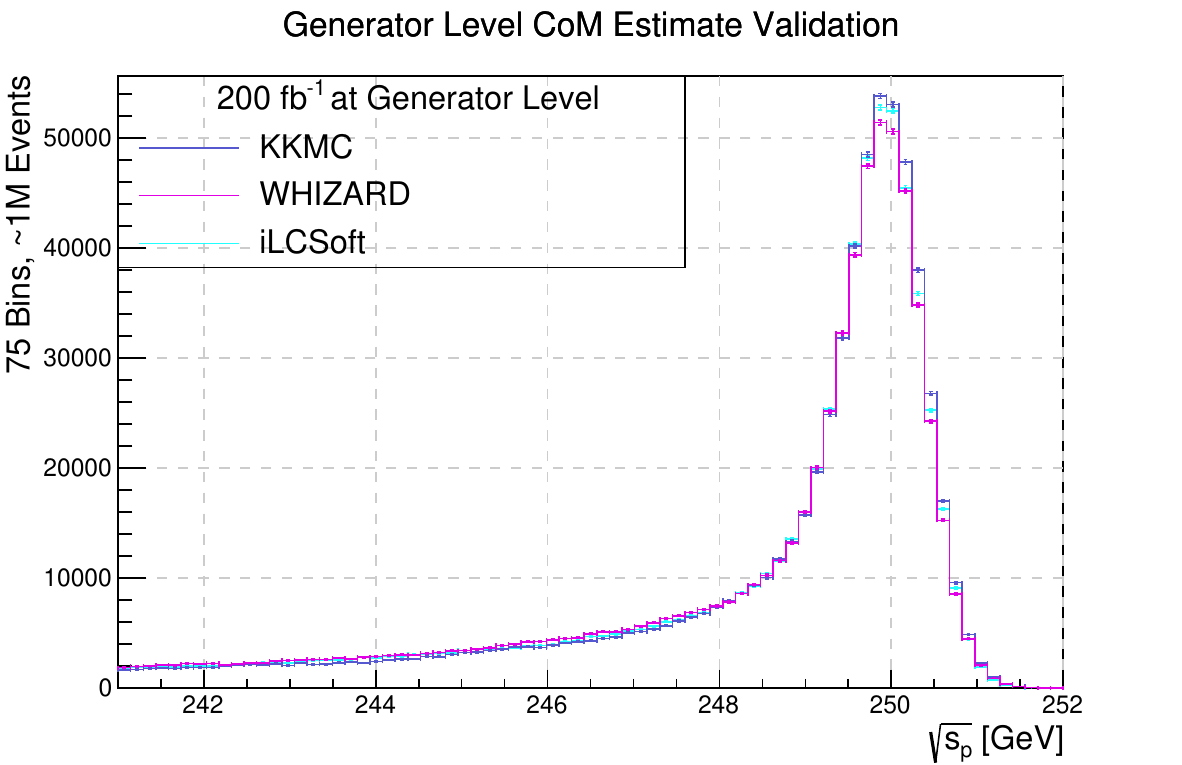}
\caption[]{\small \sl 
Each of the three DiMuon datasets no net initial-state polarization with 
an integrated luminosity of 200 $fb^{-1}$. GP2WHIZ included corrections up to two ISR photons and three additional as can be seen in~\ref{sec:A1}.
}
\label{fig:GenCheck}
\end{figure}

To further investigate discrepancies we look at the equal width bin distributions at generator level and detector level. As seen in Fig \ref{fig:GenCheck}, we observe that GP2WHIZ has the most spread into the tail region while GP2KKMC has the least.

This is plausible as KKMC is considered to be next to leading order while WHIZARD, as used in GP2WHIZ, includes at least two ISR photons and three additional photons. Potentially including more explicit photon effects. Thus KKMC may have less radiative effects, and thus less energy losing spread, as WHIZARD as used in GP2WHIZ. We provide an example WHIZARD Sindarin file in \ref{sec:A1} for the sake of reproducibility. We have chosen to compute the higher order matrix elements in said Sindarin file. For iLCSoft the WHIZARD settings and outcome were not reproducible. Given Fig \ref{fig:GenCheck} one might presume that iLCSoft's implementation has one or two photons in addition to ISR.

\begin{figure}[!h]
\centering
\includegraphics[width=0.75\textwidth]{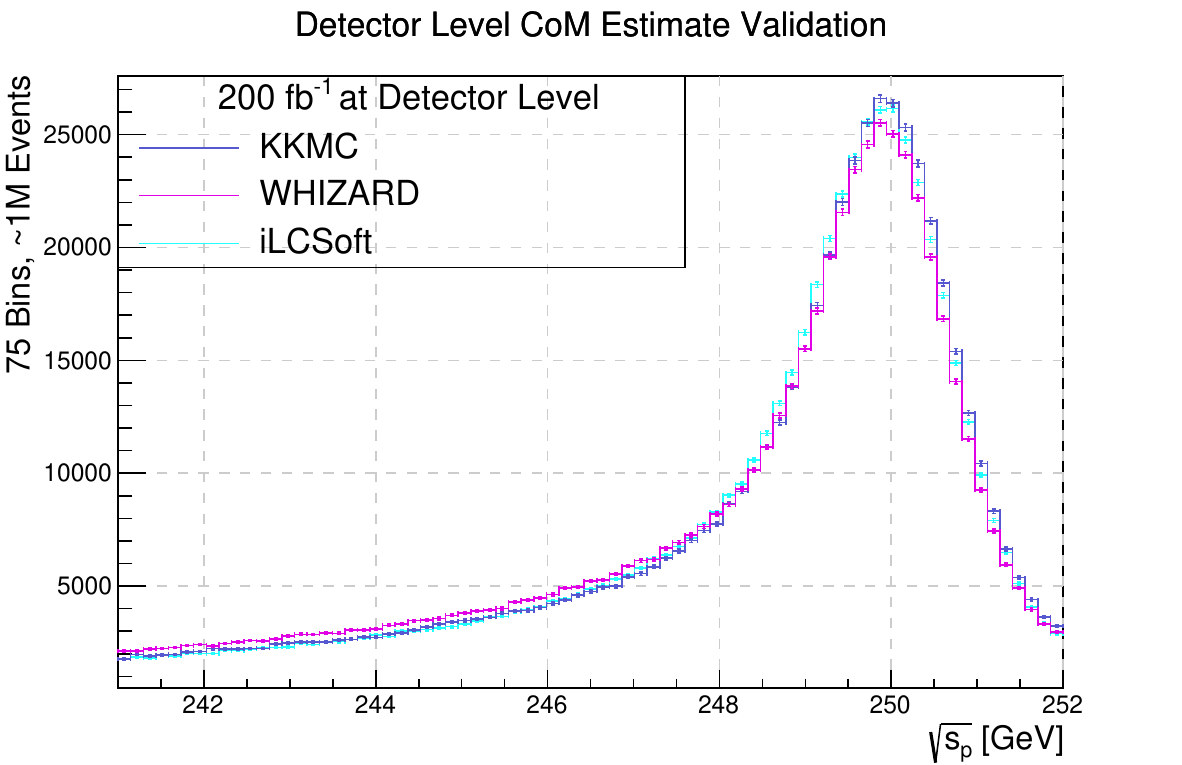}
\caption[]{\small \sl 
Each of the three datasets generated 200~$\invfb$ net unpolarized DiMuons. The tracker smearing was approximated using a parametric smear in tracked $p_T$ as explained in other work \cite{Wilson}. There are slight deviations but this is expected due to the approximated tracker performance.
}
\label{fig:DETCheck}
\end{figure}

We repeat for detector level data, as seen in Fig \ref{fig:DETCheck}. We observe the same trends as before with respect to the spread into the tail region.

To further investigate the datasets we use $\chi^2/NDoF$ for an equipartition bin, ``equibinned'', histogram. This is done so that there are no low statistic bins. We normalize the bins to 1 and use iLCSoft as the reference for creating the bin edges and normalization. We estimate the systematic uncertainty with the assumption that the systematic uncertainty, along with the statistical uncertainty, yields a reduced $\chi^2$ of unity. The evaluation is broken into two ranges, the peak $[248,252]$ and including the tail $[241,252]$. This is done for both generator level and detector level. 

\begin{figure}[!h]
\centering
\includegraphics[width=0.75\textwidth]{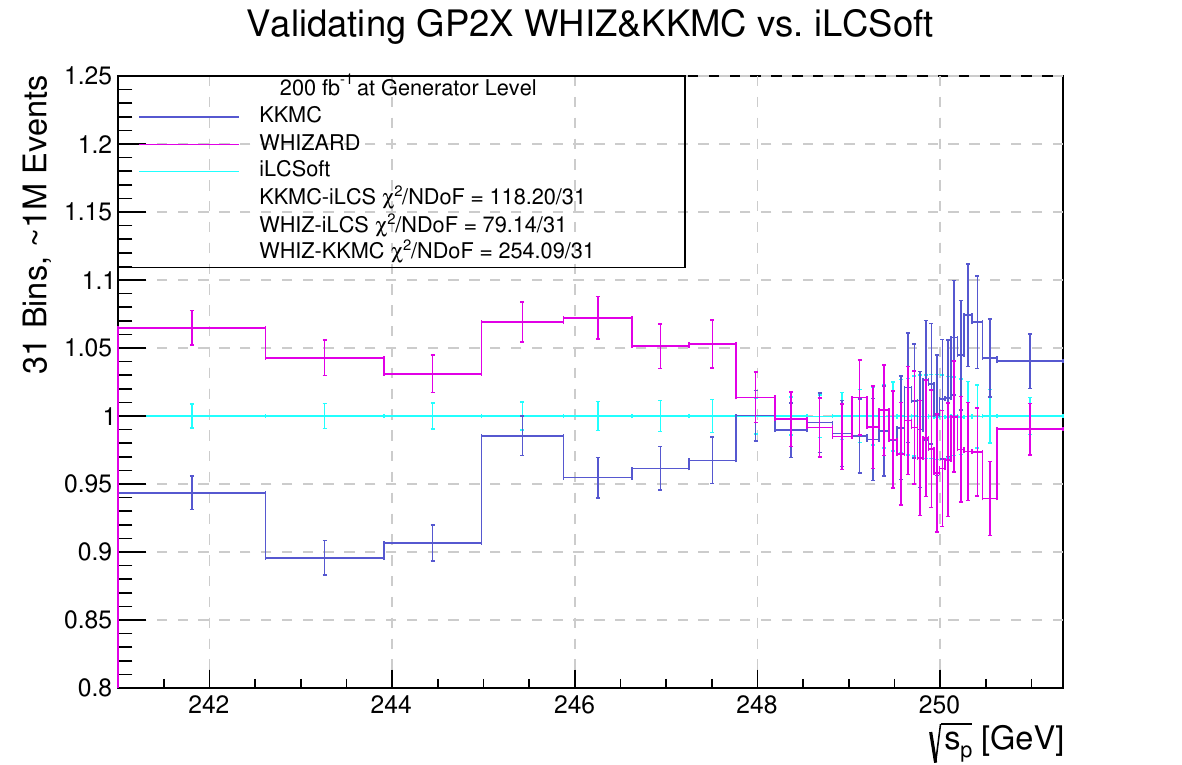}
\caption[]{\small \sl 
Generated using iLCSoft, GP2KKMC and GP2WHIZ with unpolarized initial beam particles and final state DiMuons.  
}
\label{fig:GenEqui}
\end{figure}

As seen in Fig \ref{fig:GenEqui}, generator level finds unsatisfactory $\chi^2$ over the entire region. All can be rejected as being similar at 1\% confidence. Systematic uncertainties correspond to this are 0.024 for GP2WHIZ compared to iLCSoft, 0.031 for GP2KKMC compared to iLCSoft and 0.060 for GP2WHIZ compared to GP2KKMC.

\begin{figure}[!h]
\centering
\includegraphics[width=0.75\textwidth]{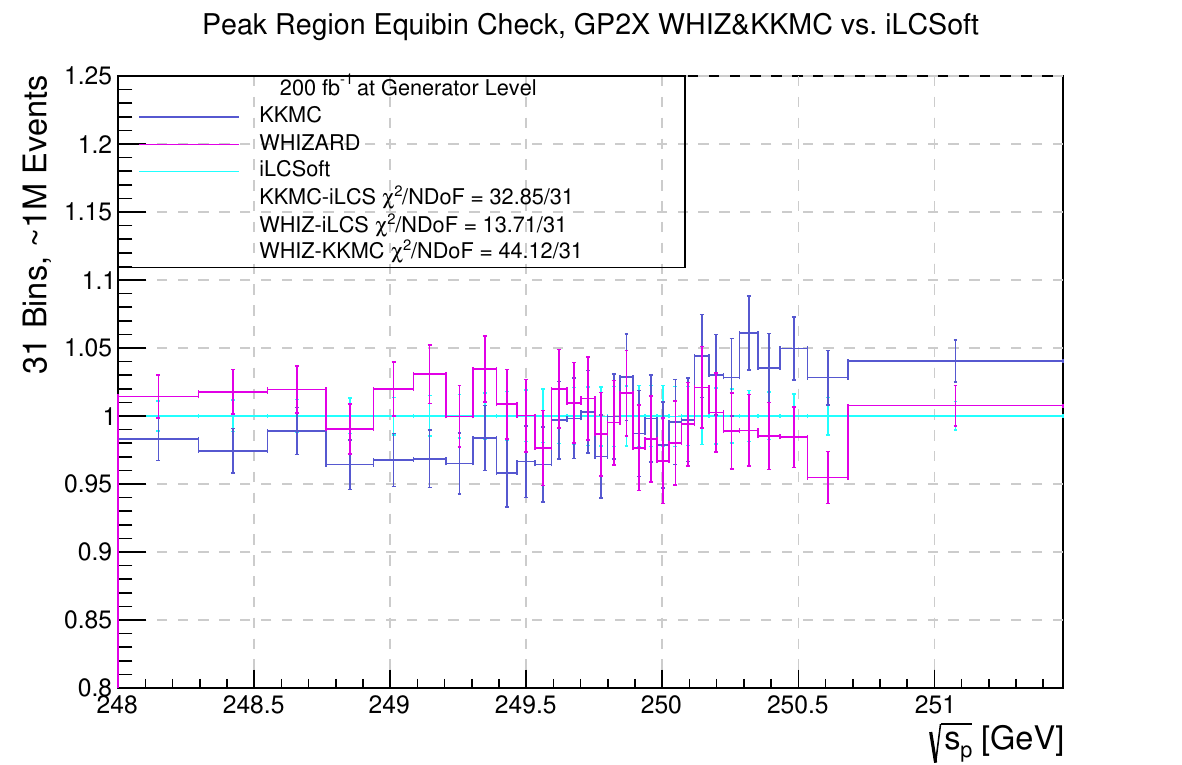}
\caption[]{\small \sl 
Normalized equibinned histogram of $\sqrt{s_{p}}$ using generator level data of DiMuons generated using GP2WHIZ, GP2KKMC and iLCSoft.
}
\label{fig:GenPeak}
\end{figure}

As seen in Fig \ref{fig:GenPeak}, we find better performance when restricting to the peak region, where the majority of data is. In the peak region with iLCSoft compared to both GP2WHIZ and GP2KKMC, both can be accepted at 1\% confidence. At the same confidence GP2KKMC compared to GP2WHIZ can be rejected. Given these generator level results it is clear that the tail region dominates the differences between the simulations. More work on verifying simulation performance in the tail region needs to be done to resolve these issues. Particularly we note that this cannot be explained simply as a GP2X issue since the confidence in acceptance changes significantly between GP2KKMC and GP2WHIZ. As such luminosity simulation, done with GUINEAPIG in GP2X and approximated using CIRCE2 in iLCSoft, should be investigated. The treatment of higher order, radiative, physics effects should also be investigated in the event generators.

\begin{figure}[!h]
\centering
\includegraphics[width=0.75\textwidth]{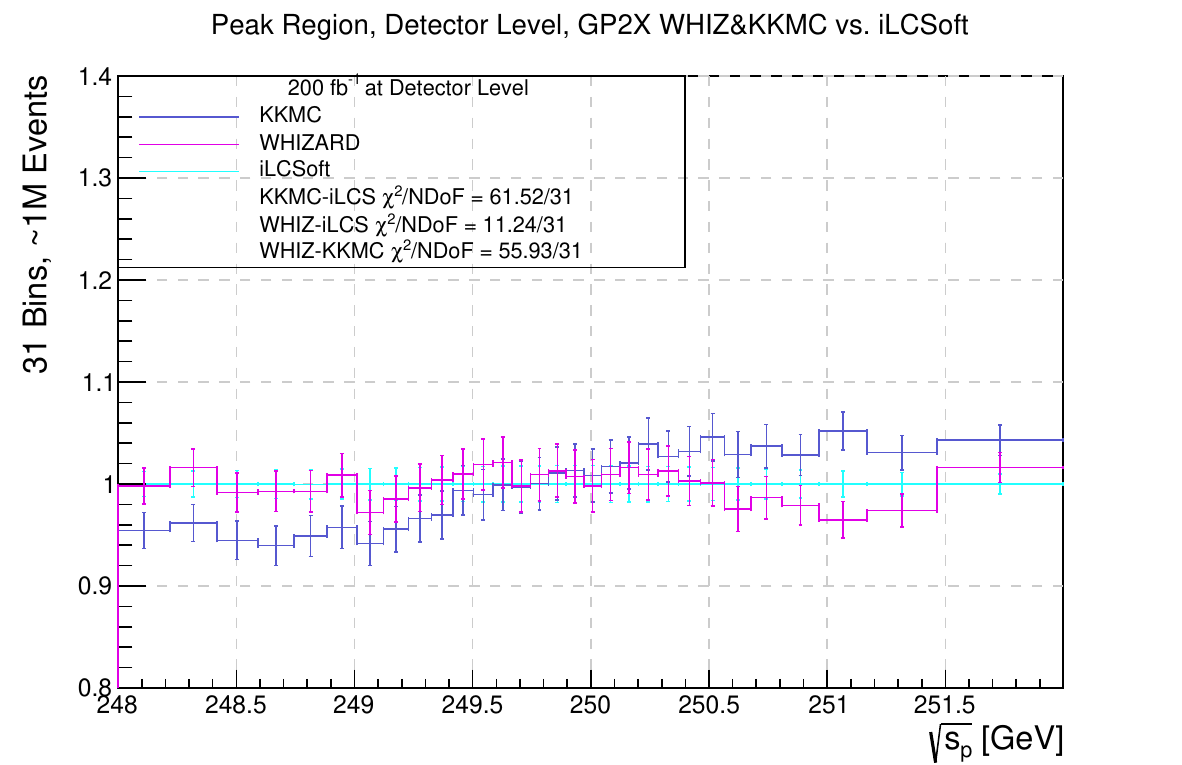}
\caption[]{\small \sl 
Normalized equibinned histogram $\sqrt{s_{p}}$ using detector level data of DiMuons generated using GP2WHIZ, GP2KKMC and iLCSoft.
}
\label{fig:DetTail}
\end{figure}

For detector level we find similar unsatisfactory $\chi^2$ over the entire region. With systematic uncertainties corresponding to 0.065 for iLCSoft with GP2WHIZ, 0.019 for iLCSoft with GP2KKMC, and 0.064 for GP2WHIZ with GP2KKMC. Note that the systematic uncertainty is smaller for GP2KKMC for reasons seen in Fig \ref{fig:DetTail}.

We observed that GP2KKMC undershoots iLCSoft below the peak and overshoots above the peak. Therefore the net deviation is smaller for GP2KKMC. This can be explained as being dominantly from the spread to the tail trend as seen in Fig \ref{fig:GenCheck} and Fig \ref{fig:DETCheck}. When restricting to the peak region we find that GP2WHIZ and iLCSoft can be accepted as being similar at 1\% confidence while the other permutations can be rejected. Since this is done at detector level the increased disagreement is likely from the difference in detector treatment between iLCSoft and GP2X. Still, in the peak region, GP2WHIZ is similar to iLCSoft to 0.31\% confidence at generator level and 0.04\% confidence at detector level.

To resolve the discrepancies in the iLCSoft and GP2WHIZ tail region we conducted tests with different amounts of added photons in the WHIZARD input file. Our initial tests were done with up to three photons. Restricting to one photon yielded results that can be rejected above 1\% confidence. Using two photons finds satisfactory results. Evaluating $\chi^2$ over the entire region at generator level, as done in Fig \ref{fig:GenEqui}, finds $\chi^2 / NDoF$ of 37.5 / 31 and a p-value of 0.196. Thus GP2WHIZ, with two photons and 200 $\invfb$ of data, compared to iLCSoft cannot be rejected at generator level. It is unclear if this discrepancy is because of iLCSoft possibly truncating at two photons or using some other method to handle higher order physics.

We now seek to estimate the performance of the parametric detector model used by GP2X. We first assume that the detector model effects dominate the differences between the generator level and detector level. We can then evaluate the systematic uncertainties at both levels. We choose to restrict this to the peak region so as to avoid issues with the tail spread. We also choose to restrict to only using GP2WHIZ since it matched iLCSoft best. From this we find that going from generator to detector level increased the normalized systematic uncertainty by 0.0048. This is to say, the discrepancies between GP2WHIZ detector modeling and iLCSoft detector modeling, within the peak region, accounts for a 0.48\% change. Since iLCSoft includes full ILD detector simulation and reconstruction we expect deviations. The detector model used by GP2X's parametric approximation could be improved by including more parameters and more fit regions. Considering these tests it is plausible to state that only GP2WHIZ, and not GP2KKMC, is operating consistent with iLCSoft.

\pdfoutput=1
\section{Beta Convolution Method}\label{sec:beta}

\begin{figure}[!h]
\centering
\includegraphics[width=0.85\textwidth]{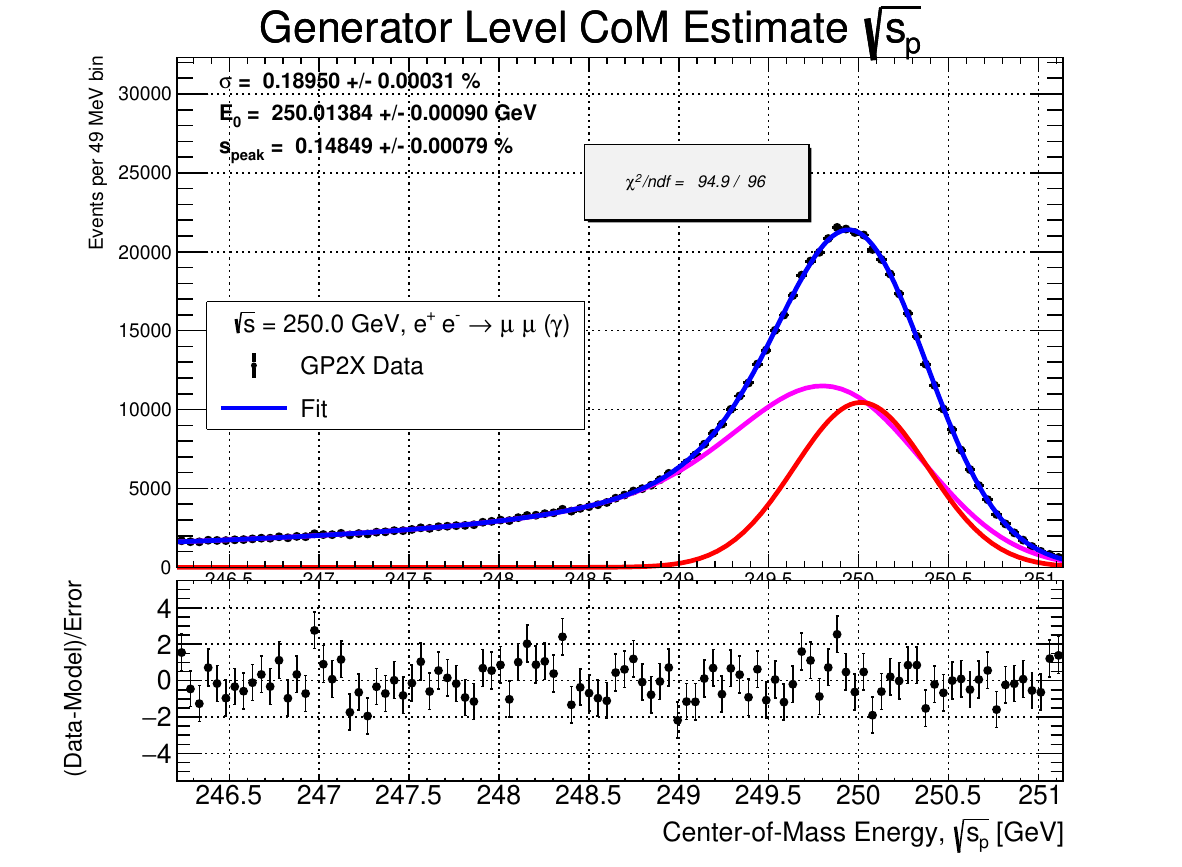}
\caption[]{\small \sl 
KKMC and GP2X have been used to generate the 200 $\invfb$ generator level dataset used in this plot. Fitting was done using the beta convolution with Gaussian added method and RooFit. The red line indicates the pure Gaussian contribution of Equation~\ref{eqn:conv2} while the magenta line indicates the Beta convolved contribution. The $\chi^{2}$ indicates a plausible fit and the $s_{peak}$ value is within 1.9\% of the design beam energy spread.
}
\label{fig:beta}
\end{figure}

Using the momentum based estimator of center of mass energy as discussed in \cite{BMadison}, $\sqrt{s_p}$, we attempt to fit for the center of mass energy at generator level. We would like to include generator level effects in our fit, particularly ones of beamstrahlung, beam energy spread, and the convolution of the physics cross-section with the luminosity spectrum. Previous work on one dimensional fitting of the luminosity spectrum absent beam energy spread effects has been done using beta distributions in CIRCE \cite{Ohl:1996fi}. In recent developments a model using the sum of a beta distribution convolved with a Gaussian and an additional Gaussian distribution has been used \cite{Wilson}. This distribution has five fit parameters, $\alpha , \beta$ for the beta distribution, 
$\mu , \sigma$ for the Gaussian distribution, and $f_{peak}$ for the fraction of the two in the sum. Initial use of this Gaussian Peak and Convolved Beta Tail model was not satisfactory at fitting all the permutations of $\sqrt{s_p}$ and $E_{\pm}^C$ distributions.
The $\sqrt{s_p}$ fits also failed to recover the correct intrinsic beam energy spread and instead were recovering values that were roughly 1.1 times too large. To address this the convolution was rewritten to allow for the two Gaussians in the convolution to still share their means but have different variances. Such that the probability density in terms of energy fraction $x = \frac{\sqrt{s}}{\sqrt{s_{nom.}}}$ is
\begin{multline}
    p(\xp; \alpha, \beta, \sigma, s_{peak}, E_{0}, f_{peak}) = \int_{0}^{1}
     f_{peak} \delta(x - 1) G(\xp-x;s_{peak}) + ...\\ 
     ... + [\frac{(1-f_{peak})}{B(\alpha, \beta)} 
     x^{\alpha-1}(1-x)^{\beta-1}
     ] \; G(\xp-x; \; \sigma)  dx \;  
    \label{eqn:conv}
\end{multline}

where we are convolving to the new energy fraction of $x' = \frac{\sqrt{s'}}{\sqrt{s_{nom.}}}$. Eqn.~\ref{eqn:conv} can be simplified 

\begin{multline}
    p(\xp; \alpha, \beta,[H] \sigma, s_{peak}, E_{0}, f_{peak}) = 
     f_{peak} G(\xp-1;s_{peak}) + ...\\ 
     ... + \int_{0}^{1} [\frac{(1-f_{peak})}{B(\alpha, \beta)} 
     x^{\alpha-1}(1-x)^{\beta-1}
     ] \; G(\xp-x; \; \sigma)  dx \;  
    \label{eqn:conv2}
\end{multline}

so the delta function is removed and now the sum is outside the integral. As a part of this work this model was implemented into RooFit~\cite{RooFit}. With this six parameter variation of the model the fit quality was improved and the intrinsic beam energy spreads, as seen in Table~\ref{tab:beta}, measured values similar to the nominal values. An example of a generator level fit can be seen in Figure~\ref{fig:beta}

\section{Fit Results}

Results for the center-of-mass energy scale at generator and detector level are now presented for various combinations of 
luminosity spectrum model and physics event generator. We also include fits of the 
distribution of inferred lab-frame energies of the colliding electron and positron beams using the estimator discussed in~\cite{Wilson}.
When including detector level effects the fits became undesirable. For this reason the detector level results quoted here will be derived as follows. Starting with the generator level precision the precision shall be scaled by 
\begin{equation}
\zeta_{GEN2TOT} = \frac{\sigma_{TOT}}{\sigma_{GEN}}
\end{equation}
the ratio of the Gaussian spread of the generator level data and total spread, i.e detector level data. These are both obtained by fitting the Gaussian peak and right portion of the peak of the data. Which is where the dominant effect is the Gaussian beam energy spread. For the total spread and generator spread at the detector level and generator level respectively. Thus, if the generator level precision is 2 ppm and $\zeta_{GEN2DET}$ is 2.0 then the detector level precision will be quoted as 4 ppm. For this study it was found that the ILD tracker, for $\sqrt{s_{p}}$, had $\zeta_{GEN2DET}\approx$~1.76 while the ECAL had $\zeta_{GEN2DET}\approx$~9.34. In doing so we are assuming a similar quality of fit is possible at the detector level as has been done at the generator level. For simplicity the symbol $\epsilon_{\mu}$ is used to refer to the precision of the mean of the fits. Results, as seen in Table \ref{tab:beta}, indicate that an energy precision goal near 1.0 ppm is plausible for the 100 $fb^{-1}$ dataset. Extrapolating to a larger, 2000 $fb^{-1}$, dataset one could also find an energy precision goal near 0.25 ppm to be plausible. Still, this beta convolution method is approaching but not meeting these goals. We also observe two peculiarities. First, that the electron beam, which is broader in spread, typically is resolved with more precision than the positron beam. Second we observe that KKMC and WHIZARD do not have similar results while BHWIDE and WHIZARD share nearly identical results. We are not sure the reasons for these observations and leave speculation up to the reader.

\begin{table}[!h]
\begin{tabular}{|ccccc|}
\hline
\multicolumn{5}{|c|}{Using ILC250 + ILD, 100 $fb^{-1}$, Beta Convolution + Gaussian}                                                                                                                                                                 \\ \hline
\multicolumn{1}{|c|}{Metric}                          & \multicolumn{1}{c|}{Generator $\epsilon_{\mu}$ (ppm)} & \multicolumn{1}{c|}{Detector $\epsilon_{\mu}$ (ppm)} & \multicolumn{1}{c|}{Fit $\sigma$ [GeV]}   & $\Delta E_{0}$ (ppm) \\ \hline
\multicolumn{1}{|c|}{iLCSoft $\sqrt{s_{p}}$}             & \multicolumn{1}{c|}{4.8}                                & \multicolumn{1}{c|}{11}                & \multicolumn{1}{c|}{0.38 $\pm$ 0.05} & 30                   \\ \hline
\multicolumn{1}{|c|}{iLCSoft $E_-^C$}                & \multicolumn{1}{c|}{7.9}                                & \multicolumn{1}{c|}{14}                & \multicolumn{1}{c|}{0.24 $\pm$ 0.01}               & 40                    \\ \hline
\multicolumn{1}{|c|}{iLCSoft $E_+^C$}                & \multicolumn{1}{c|}{6.5}                                & \multicolumn{1}{c|}{14}                & \multicolumn{1}{c|}{0.19 $\pm$ 0.05}               & 85                    \\ \hline

\multicolumn{1}{|c|}{KKMC $\sqrt{s_{p}}$}             & \multicolumn{1}{c|}{4.8}                                & \multicolumn{1}{c|}{10}                & \multicolumn{1}{c|}{0.37 $\pm$ 0.01} & 55                    \\ \hline
\multicolumn{1}{|c|}{KKMC $E_-^C$}                & \multicolumn{1}{c|}{7.5}                                & \multicolumn{1}{c|}{13}                & \multicolumn{1}{c|}{0.24 $\pm$ 0.01}               & 87                    \\ \hline
\multicolumn{1}{|c|}{KKMC $E_+^C$}                & \multicolumn{1}{c|}{6.0}                                & \multicolumn{1}{c|}{12}                & \multicolumn{1}{c|}{0.19 $\pm$ 0.01}               & 99                    \\ \hline
\multicolumn{1}{|c|}{WHIZARD $\mu\mu$ $\sqrt{s_{p}}$} & \multicolumn{1}{c|}{4.4}                                & \multicolumn{1}{c|}{9.3}                & \multicolumn{1}{c|}{0.38 $\pm$ 0.01} & -12          \\ \hline
\multicolumn{1}{|c|}{WHIZARD $\mu\mu$ $E_-^C$}    & \multicolumn{1}{c|}{8.1}                                 & \multicolumn{1}{c|}{14}                & \multicolumn{1}{c|}{0.24 $\pm$ 0.02}               & 56                    \\ \hline
\multicolumn{1}{|c|}{WHIZARD $\mu\mu$ $E_+^C$}    & \multicolumn{1}{c|}{6.6}                                & \multicolumn{1}{c|}{13}                & \multicolumn{1}{c|}{0.19 $\pm$ 0.01}               & 11                    \\
\hline
\multicolumn{1}{|c|}{WHIZARD $ee$ $\sqrt{s_{p}}$}     & \multicolumn{1}{c|}{0.22}                                  & \multicolumn{1}{c|}{2.1}                & \multicolumn{1}{c|}{0.36 $\pm$ 0.02}   & 290          \\ \hline
\multicolumn{1}{|c|}{WHIZARD $ee$ $E_-^C$}        & \multicolumn{1}{c|}{0.24}                                  & \multicolumn{1}{c|}{1.4}                 & \multicolumn{1}{c|}{0.24 $\pm$ 0.01}               & -25                    \\ \hline
\multicolumn{1}{|c|}{WHIZARD $ee$ $E_+^C$}        & \multicolumn{1}{c|}{0.19}                                  & \multicolumn{1}{c|}{1.3}                & \multicolumn{1}{c|}{0.19 $\pm$ 0.01}               & -34                    \\ \hline
\multicolumn{1}{|c|}{BHWIDE $\sqrt{s_{p}}$}           & \multicolumn{1}{c|}{0.18}                                  & \multicolumn{1}{c|}{1.7}                & \multicolumn{1}{c|}{0.37 $\pm$ 0.01}   & 20                  \\ \hline
\multicolumn{1}{|c|}{BHWIDE $E_-^C$}              & \multicolumn{1}{c|}{0.28}                                  & \multicolumn{1}{c|}{1.8}                & \multicolumn{1}{c|}{0.24 $\pm$ 0.01}               & 38                    \\ \hline
\multicolumn{1}{|c|}{BHWIDE $E_+^C$}              & \multicolumn{1}{c|}{0.22}                                  & \multicolumn{1}{c|}{1.5}                & \multicolumn{1}{c|}{0.19 $\pm$ 0.01}               & 37                   \\ \hline
\end{tabular}
\caption[]{\small \sl 
Comparison of fit energy precision and fit spread across various simulations. We investigate $\sqrt{s_{p}}$ and the beam collision energies. For $\sqrt{s_{p}}$, $E_-^C$ and $E_+^C$ the expected spreads are 0.378 GeV, 0.238 GeV and 0.188 GeV respectively. We include a column ppm deviation of the fit $E_{0}$ as compared to the nominal values to show how much deviation there is in fit energy scale across the models.\\
}
\label{tab:beta}
\end{table}

We use 
\begin{equation}
\Delta E_0 = \frac{E_0^{X} - E_0^{nom.}}{E_0^{nom.}}
\end{equation}
to gauge the deviation in the fit energy scale parameters across the various simulations. This is done using the nominal values as the reference and X being other simulations like GP2WHIZ or iLCSoft. The GP2WHIZ Bhabha simulations do the poorest in matching the energy scale. It is unsure whether this is dominated by fundamental physics, such that Bhabha scattering is dominantly t-channel, or if this is a computational issue. To resolve this issue we propose a future running of GP2BHWIDE using BHWIDE's option to run using only s-channel contributions and then another run using only t-channel contributions. The remaining simulations are similar in order of magnitude of energy scale deviation. Of the DiMuon simulations we observed that GP2WHIZ is closest to the nominal values. To remedy these energy scale deviations additional running at other center-of-mass energies will be needed. This will allow for full characterization of the energy scale deviations as a function of center-of-mass energy. Using this the data can then be recalibrated to have the correct center-of-mass energy.

\pdfoutput=1
\section{Fourier Transform Deconvolution Method}

In a dataset consisting of detector level data for DiMuons measured in a detector we may use a convolution description, such as that found in radio arrays that get images from radio signals. In such a method one could write a distribution of the Fourier transform $\sqrt{s_p}$ as
\begin{equation}
F_{gen.}(\sqrt{s_p}) = R_{det.}^{-1}(\sqrt{s_p}) F_{det.}(\sqrt{s_p}) 
\end{equation}
where we describe the generator level data, $F_{gen.}(\sqrt{s_p})$, as the detector level data, $F_{det.}$, multiplied by the inverse of the detector response function, $R_{det.}$. Since this is done using a product of Fourier transform, the eigenfunctions, this is equivalent to a deconvolution of the detector level data to generator level data. In linear algebra approaches the response function is instead a tensor describing the response and the inverse matrix is used \cite{Blobel}. The Fourier approach is advantageous as, in the limit of large numbers, it is computationally easier than computing inverse matrices. In some cases such matrices can only be approximated.

The Fourier transform method is not without issues. For a binned dataset, as is typically used in high energy physics, the Fourier transform is the Discrete Fourier Transform (DFT). As such one must choose bin sizes and the window size for the dataset, both of which are non-trivial choices that have immediate effects on the DFT quality. The window size affects the frequency bandwidth of the DFT as can be calculated using the Shannon-Nyquist sampling theorem. The bin sizes affect the smoothness of the overall binned data. When bins are too fine the statistical fluctuations between bins is treated like impulsive noise would be treated in a continuous signal. Impulsive noise introduces considerable white noise in a DFT, which, in turn, leads to increased oscillatory, harmonic, noise in the output of the DFT. This harmonic noise has been observed with previous use of inverse methods like the DFT, making them undesirable \cite{Blobel}. This outcome was also reproduced in the work done here. This noise, as well as the issues of methods that may overfit or underfit and thus require regularization, are key issues in any deconvolution method.  Especially in precision measurements, any harmonic noise would deteriorate the fit quality and thus the precision of the measurement.

\begin{figure}[!h]
\centering
\includegraphics[width=0.75\textwidth]{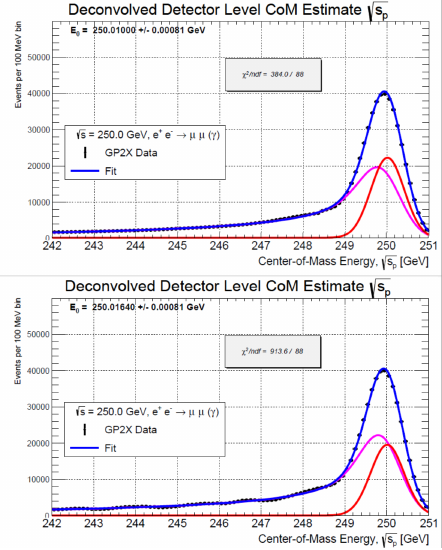}
\caption[]{\small \sl 
Using GP2WHIZ and DiMuons the detector level data for $\sqrt{s_p}$ was deconvolved to the generator level data using the Fourier transform method explained in this work. As in Figure~\ref{fig:beta}, the red line indicates the pure Gaussian contribution of Equation~\ref{eqn:conv2} while the magenta line indicates the Beta convolved contribution. The above plot has a simple implementation of the Savitzky-Golay filter for a 0.5 GeV window and 2rd order polynomial. The bottom plot is without the filter. A significant improvement in fit quality is observed.
}
\label{fig:svgtest}
\end{figure}

In digital signal processing the Savitzky-Golay filter is used to remedy data of harmonic noise, especially oscillations that are higher frequency than the data \cite{svg}. The Savitzky-Golay filter is done using successive least squares fits of a given polynomial order to the data within a window that travels across the data with a given window size. At lowest order it is equivalent to a moving average. In this study we observed oscillations in the output of the DFT of $\sqrt{s_p}$ that correspond to roughly 1 GeV in period. Thus a window length of roughly 0.5 GeV was used so that, on average, the oscillations averaged to zero across the window. We do not anticipate losing physically relevant shape information as the distribution does not have non-polynomial features smaller than the scale of the 0.5 GeV window. As can be seen in Figure \ref{fig:svgtest}, the Savitzky-Golay filter improved the fit $\chi^2$ from 914 to 384 for 88 degrees of freedom. 

Considering the presence of additional harmonic noise after a single use of the Savitzky-Golay filter, this method can be improved.

After the filter is applied we have a deconvolved generator level dataset that, with less noise, may be desirable. This deconvolved distribution is fit to the same six parameter fit used in Section~\ref{sec:beta}. Results indicate that this function has similar fit quality to what was observed in Section \ref{sec:beta}. Particularly we use the GP2WHIZ DiMuon $\sqrt{s_p}$ fit as an example. As a reference, the ratio of spread between generator level and detector level quoted in Section \ref{sec:beta} was 1.76. An ideal deconvolution would be comparable, or better, than this ratio. Using the Fourier transform deconvolution and Savitzky-Golay filter approach we find that the fitted ratio was 2.38. This test would then expect the current $\sqrt{s_p}$ fit quality to be systematics dominated. A compilation of fit results for GP2WHIZ using DiMuons and Bhabhas can be found in Table \ref{tab:unf}. We observe that, as the preliminary test showed, the results are dominated by systematic uncertainties. Additional work will need to be done to precisely identify and reduce the sources of systematic uncertainty.

\begin{table}[]
\begin{tabular}{|cccc|}
\hline
\multicolumn{4}{|c|}{Using GP2WHIZ, ILC250 + ILD, 100 $\invfb$ data, 1600 $\invfb$ reference, Deconvolution Fit}                                                                                                                         \\ \hline
\multicolumn{1}{|c|}{Metric}                          & \multicolumn{1}{c|}{Total Unc. (ppm)} & \multicolumn{1}{c|}{Sys. Unc. (ppm)} & $\Delta E_0$ (ppm) \\ \hline
\multicolumn{1}{|c|}{$\mu\mu$ $\sqrt{s_{p}}$} & \multicolumn{1}{c|}{6.9}                                        & \multicolumn{1}{c|}{6.0}                & 40        \\ \hline
\multicolumn{1}{|c|}{$\mu\mu$ $E_-^C$}    & \multicolumn{1}{c|}{11}                                        & \multicolumn{1}{c|}{6.3}                & 162                    \\ \hline
\multicolumn{1}{|c|}{$\mu\mu$ $E_+^C$}    & \multicolumn{1}{c|}{15}                                        & \multicolumn{1}{c|}{7.5}                & -307                    \\ \hline
\multicolumn{1}{|c|}{$ee$ $\sqrt{s_{p}}$}     & \multicolumn{1}{c|}{0.28}                                        & \multicolumn{1}{c|}{0.25}                & 313          \\ \hline
\multicolumn{1}{|c|}{$ee$ $E_-^C$}        & \multicolumn{1}{c|}{0.35}                                        & \multicolumn{1}{c|}{0.26}                 &  51                  \\ \hline
\multicolumn{1}{|c|}{$ee$ $E_+^C$}        & \multicolumn{1}{c|}{0.40}                                        & \multicolumn{1}{c|}{0.30}                &  121                   \\ \hline
\end{tabular}
\caption[]{\small \sl 
This table was made to be similar to Table~\ref{tab:beta}. The systematic uncertainty is estimated assuming that the total uncertainty is the quadratic sum of statistical and systematic. Data was fit after using the Fourier transform deconvolution method and being filtered once with a Savitzky-Golay filter.
}
\label{tab:unf}
\end{table}

We observe that, similar to the previous results in Table~\ref{tab:beta}, there are significant energy scale deviations. These deviations seem to have been worsened by the deconvolution method. Thus, as with the previous results, there will need to be additional runs at other center-of-mass energy values to calibrate this energy scale deviation.

\pdfoutput=1
\section{Conclusions}
\label{sec:concl}
We have developed software and methods for future $\ee$ Higgs factories. Using channels like $\mumu$ and $\ee$ with $\sqrt{s_{p}}$ and software like GP2X and the Fourier transform deconvolution fit method there has been significant progress towards the sub 1 ppm energy precision goals of these future experiments. These have been checked against work of other research using multiple methods. Therefore the progress is plausibly robust. This work needs to be expanded to include more higher order processes and to have some evaluation and treatment of backgrounds. All polarizations need to be developed and presented. Additional center-of-mass energy values need to be investigated for both calibrating the energy scale deviations and for characterizing the other possible center-of-mass energies that will be used in proposed colliders. To test the robustness of the fits and deconvolution method work using the larger datasets, of 2 $ab^{-1}$ integrated luminosity, is needed. This work could also be extended to other colliders or detectors. We should also include calibration of the tracker performance, and the ECAL performance, so that these results are more robust. For the deconvolution fitting method the systematic uncertainty must be reduced by better implementation of the Savitzky-Golay filter or similar methods. There are also other deconvolution methods that may be more robust to noise and regularization issues. The deconvolution method also has more utilities, such as deconvolving the accelerator luminosity from detector data. As such this work holds promise for continued progress.

\section*{Acknowledgments}
We acknowledge Graham Wilson for discussions, previous work, work on documenting the rescaling method used here and the derivation of the beta convolution fit.
This work is partially supported by the US National Science Foundation under award NSF 2013007. This work benefited from use of the HPC facilities operated by the Center for Research Computing at the University of Kansas. 



\pdfoutput=1
\newpage
\appendix

\section{Example WHIZARD Sindarin File}\label{sec:A1}
\text{ }\\
\verb|model = SM_Higgs_CKM|\\
\verb|mmu = 0.105658|\\
\verb|me = 0.000511|\\
\verb|mW = 80.419|\\
\verb|mZ = 91.1876|\\
\text{ }\\
\verb|polarized "e1","E1","e2","E2"|\\
\text{ }\\
\verb|sample_format = lhef|\\
\text{ }\\
\verb|process mumu = "e1", "E1" => ("e2","E2") + |\\
\verb|              ("e2", "E2", "A") + ("e2", "E2", "A", "A") + | \\
\verb|              ("e2", "E2", "A", "A", "A")| \\
\text{ }\\
\verb|cuts = all 5 degree < Theta < 175 degree [e2] and|\\
\verb|       all 5 degree < Theta < 175 degree [E2]|\\
\text{ }\\
\verb|beams = e1, E1 => isr, isr|\\
\verb|#beams_momentum = 0.5 * VCOM , 0.5 * VCOM|\\
\verb|#beams_theta = 0.802 degree , 0.802 degree|\\
\verb|#!isr_order         = 3|\\
\verb|?isr_handler       = true|\\
\verb|$isr_handler_mode  = "recoil"|\\
\verb|isr_alpha          = 0.0072993|\\
\verb|isr_mass           = 0.000511|\\
\verb|beams_pol_density = @(+1), @(-1)|\\
\verb|beams_pol_fraction = 80%, 30%|\\
\text{ }\\
\verb|n_events = VNEV|\\
\text{ }\\
\verb|compile|\\
\text{ }\\
\verb|sqrts = VCOM|\\
\verb|integrate (mumu) { iterations = 20:10000 }|\\
\text{ }\\
\verb|simulate (mumu)|\\
\text{ }\\
\verb|write_analysis|\\
\verb|compile_analysis { $out_file = "mumu.dat" }|\\

\end{document}